\documentclass[12pt]{iopart}

\usepackage{iopams} 
\usepackage{graphicx}
\usepackage{dcolumn}
\usepackage{bm}
\usepackage{xcolor} 
\usepackage[normalem]{ulem} 
\usepackage[version=3]{mhchem} 

\usepackage[
scale=0.7, marginratio={1:1, 2:3}, 
text={7in,10in},centering,
margin=1.5in,
total={6.5in,8.75in}, top=1.2in, bottom=1.2in, left=0.9in, right=0.9in, 
height=10in,a4paper,hmargin={3cm,0.8in},
]{geometry}

\begin{document}


\title[]{Anisotropic hybridization probed by polarization dependent x-ray absorption spectroscopy in VI$_3$ van der Waals Mott ferromagnet}


\author{R Sant$^1$\footnote[1]{To whom correspondence should be addressed.}\footnote[2]{Current address: Dipartimento di Fisica, Politecnico di Milano, Piazza Leonardo da Vinci 32, I-20133 Milano, Italy.}, A De Vita$^{2,3}$, V Polewczyk$^2$, G Pierantozzi$^2$, F Mazzola$^{2,4}$, G Vinai$^2$, G van der Laan$^5$, G Panaccione$^2$, N B Brookes$^1$}
    
\address{$^1$ ESRF, The European Synchrotron, 71 Avenue des Martyrs, CS40220, 38043 Grenoble Cedex 9, France}
\address{$^2$ Istituto Officina dei Materiali (IOM)-CNR, Laboratorio TASC, in Area Science Park, S.S.14, Km 163.5, I-34149 Trieste, Italy}
\address{$^3$ Dipartimento di Fisica, Università di Milano, Via Celoria 16, I-20133 Milano, Italy}
\address{$^4$ Department of Molecular Sciences and Nanosystems, 32 Ca’ Foscari University of Venice, 30172 Venice, Italy}
\address{$^5$ Diamond Light Source, Harwell Science and Innovation Campus, Didcot, Oxfordshire OX11 0DE, UK}
    
\vspace{12pt}
\ead{roberto.sant@polimi.it}

\vspace{12pt}
\begin{indented}
\item[May 2023]
\end{indented}


\begin{abstract}
Polarization dependent x-ray absorption spectroscopy was used to study the magnetic ground state and the orbital occupation in bulk-phase VI$_3$ van der Waals crystals below and above the ferromagnetic and structural transitions. X-ray natural linear dichroism and X-ray magnetic circular dichroism spectra acquired at the V $L_{2,3}$ edges are compared against multiplet cluster calculations within the frame of the ligand field theory to quantify the intra-atomic electronic interactions at play and evaluate the effects of symmetry reduction occurring in a trigonally distorted VI$_6$ unit. We observed a non zero linear dichroism proving the presence of an anisotropic charge density distribution around the V$^{3+}$ ion due to the unbalanced hybridization between the Vanadium and the ligand states. Such hybridization acts as an effective trigonal crystal field, slightly lifting the degeneracy of the $t_{2g}^2$ ground state. However, the energy splitting associated to the distortion underestimates the experimental band gap, suggesting that the insulating ground state is stabilized by Mott correlation effects rather than via a Jahn-Teller mechanism. Our results clarify the role of the distortion in VI$_3$ and establish a benchmark for the study of the spectroscopic properties of other van der Waals halides, including emerging 2D materials with mono and few-layers thickness, whose fundamental properties might be altered by reduced dimensions and interface proximity.
\end{abstract}


\section{Introduction}

Understanding orbital and magnetic order in layered van der Waals (vdW) compounds is crucial to control the magnetic properties in few-layer thick two-dimensional (2D) materials with intrinsic (anti)ferromagnetism persistent down to the monolayer limit. Although some magnetic vdW compounds display layer-dependent behavior -- \textit{e.g.} in bilayer CrI$_3$, where the overall magnetization is suppressed due to the antiferromagnetic coupling between two adjacent layers \cite{huang} -- in most transition metal compounds the electronic and magnetic order is determined unambiguously by the electronic interactions occurring on the metal atoms, such as crystal-field and Coulomb repulsion \cite{zaanen,degroot2}. This justifies the use of multiplet cluster models \cite{degroot1} to describe theoretically the magnetic behavior of vdW $3d$ transition metal ferromagnets. Experiments carried out to validate the theory help to establish the benchmarks in bulk compounds before passing on to investigate the same materials in the 2D regime, especially when the orbital order is altered by extrinsic factors such as the substrate or the proximity of another 2D layer \cite{bora,zhao,zhangVSe2,vinai}.

\begin{figure*}[ht]
    \centering
    \includegraphics[scale=0.9]{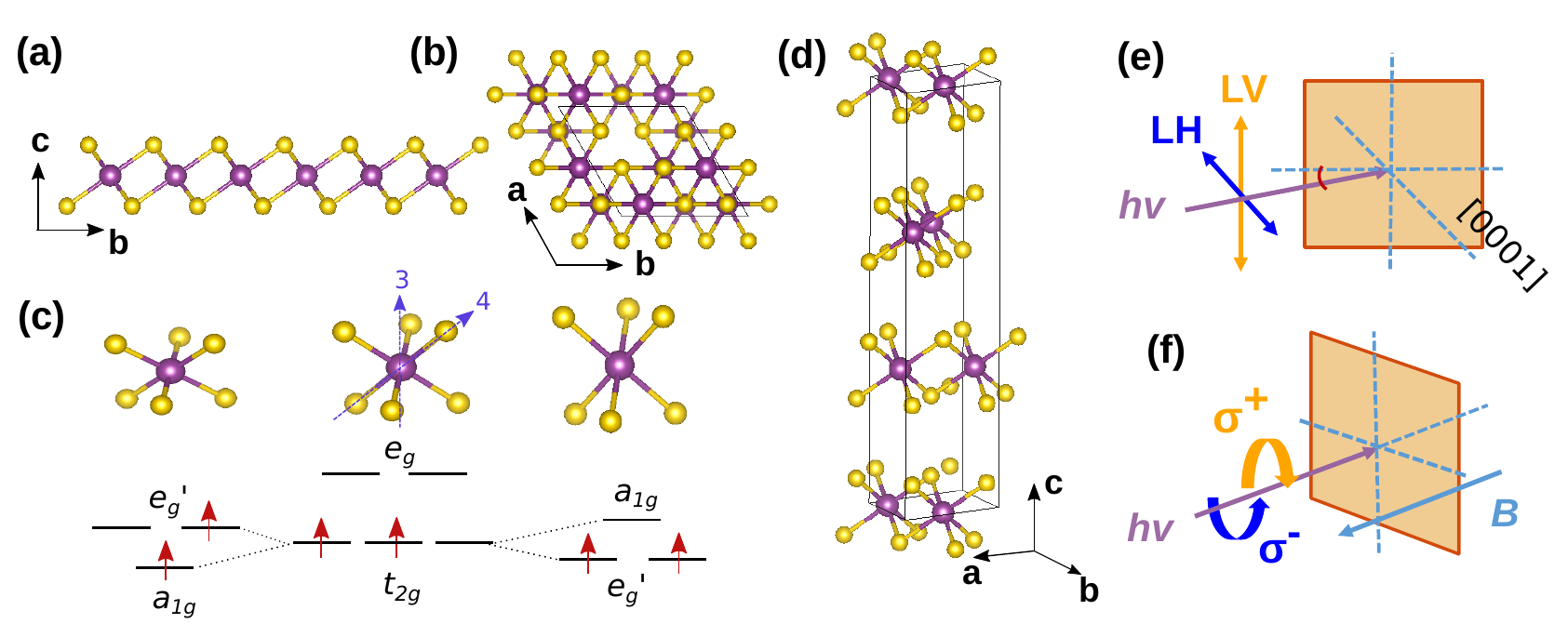}
    \caption{\textbf{VI$_3$ crystal structure and orbital order.} Side (a) and top (b) view of the hard spheres representation of a single layer VI$_3$. (c) Perspective views of the VI$_6$ cluster unit and the relative $3d$ energy levels splitting due to a trigonal Jahn-Teller deformation (compression and elongation occur along the trigonal, \textit{i.e.} 3-fold, axis). (d) Perspective view of the VI$_3$ low temperature monoclinic unit cell; the $c$ axis is along the [0001] direction perpendicular to the vdW planes. In all the illustrations, V and I atoms are depicted by purple and yellow spheres respectively.  \textbf{Principles of XNLD and XMCD techniques.} (e) In XNLD, XAS spectra are acquired with horizontal (blue) and vertical (orange) linearly polarized x-rays and in absence of any magnetic field; in our experiment the x-ray beam impings on the sample at 20$^{\circ}$ grazing incidence. (f) In a XMCD experiment, left- (blue) and right- (orange) circularly polarized x-rays while the magnetic field $B$ is applied parallel to the x-ray beam; in our experiment the x-ray beam is normal to the sample.}
    \label{fig:sketches}
\end{figure*}

While many attempts are focused on preparing few-layer-thick ferromagnetic materials within heterostructures and device prototypes in view of technological applications, many fundamental questions still remain unanswered concerning the electronic and magnetic behavior of some vdW compounds in bulk form. The recent debate about the half-metallic versus Mott insulating nature of trivalent vanadium iodide (VI$_3$) is a current example arising from a knowledge gap around its electronic ground state \cite{he, long, tian, nguyen}. VI$_3$ is a vdW transition metal halide with MX$_3$ formula (M = transition metal and X = halogen) that recently received much attention by virtue of its outstanding ferromagnetism, similarly to its sister compound CrI$_3$ \cite{mcguire}. Typically, its layer unit structure is described by a V$^{3+}$ cation occupying the center of three honeycomb lattice planes and surrounded by six I$^-$ anions arranged with octahedral coordination and located within the two external planes (Figure \ref{fig:sketches}a-c) \cite{dolevzal}. Just as CrI$_3$, VI$_3$ is an Ising ferromagnet showing long range ferromagnetic order below 50 K with out-of-plane directed magnetic moments carried by the metal atoms \cite{tian, zhao, kong}. It possesses a structural phase transition at 79 K, which transforms the overall crystal symmetry from trigonal to monoclinic upon cooling \cite{tian,dolevzal}. This phase transition is described as a change in the stacking order of the layer planes within the unit cell (Figure \ref{fig:sketches}d), which does not affect the metal-iodine bonds and the local octahedral symmetry of the VI$_6$ cluster unit \cite{dolevzal}.

From the theoretical point of view, the debate firstly focused  on the metallic versus insulating nature of VI$_3$ and lately on the origin of the insulating band gap. The first studies reported that VI$_3$ has a half-metallic character \cite{he, long}, whereas subsequent ones proposed a Mott insulating nature claiming that spin-orbit coupling and correlation effects were not accounted for in the first case \cite{tian,nguyen}. Following experimental works --- performed however above the VI$_3$ Curie temperature --- confirmed the second hypothesis by determining the existence of a semiconducting bandgap: firstly Son \textit{et al.} determined, using transmission optical spectroscopy, a direct optical gap of 0.67 eV \cite{son}, subsequently angle-resolved photoemission electron spectroscopy (ARPES) experiments revealed that the electronic band structure of VI$_3$ is characterized by a larger electronic band gap of the order of 1 eV \cite{kundu,devita}. Concerning the mechanism opening such a gap, the debate is at present still open. Contrary to Cr$^{3+}$, which possesses half-filled  $t_{2g}$ states, the V$^{3+}$ ion has only two valence electrons in the three-fold degenerate $t_{2g}$ states. This latter configuration makes VI$_3$ unstable towards a trigonal Jahn-Teller (JT) distortion, which results into an elongation/compression of the VI$_6$ unit along the threefold axis and a further energy splitting of the $t_{2g}$ states into two degenerate $e_g^\prime$ levels and one $a_{1g}$ state, as illustrated in Figure \ref{fig:sketches}c \cite{park,nguyen,yang}. In spite of this picture, a more recent interpretation tends however to diminish the importance of the JT mechanism and propose instead electron correlation as the main driving force stabilizing the Mott-insulator gap \cite{nguyen}. Recent observations reported that the bandgap observed experimentally by ARPES does not correspond to the predicted Mott gap, but to the one induced by the octahedral crystal field \cite{devita}. This contrasting result was explained by a combination of two factors, \textit{i.e.} the extreme surface sensitivity of the photoemission process, capable to probe only the topmost layers of the crystal, and the electronic doping due to the sample cleaving, which fills the otherwise empty higher energy $a_{1g}$ level. This fact demonstrates the need for new investigations based on more bulk-sensitive techniques.

At the current state of the art, experimental evidence at low temperature of which mechanisms, \textit{i.e.} JT or Mott-like, stabilizes the VI$_3$ ground state, is still lacking, while the electronic ground state has not been fully addressed experimentally. To this purpose, x-ray absorption spectroscopy (XAS) is a well-established and bulk sensitive technique to determine the orbital and spin character of ground states, where the spectra can be interpreted in terms of transitions to multiplet-split final states with the help of supporting calculations.

In this article we present the x-ray natural linear dichroism (XNLD) and x-ray magnetic circular dichroism (XMCD) measurements of bulk crystals of VI$_3$ at temperatures above and below the structural (79 K) and magnetic (51 K) transitions, respectively. The dichroic absorption spectra are interpreted using multiplet cluster calculations \cite{degroot1} in order to identify the most stable electronic ground state in the ferromagnetic phase and quantify the main intra-atomic interactions such as crystal field and $dd$ (Coulomb and exchange) interactions. Using the same set of parameters we reproduced with good agreement the XAS white lines acquired with linear and circular polarizations, this fact establishing the consistency between the information extracted from the two data set. In particular, we observed a finite experimental XNLD signal, identical at 20 and 100 K, which is the result of the anisotropic charge density distribution around the V$^{3+}$ ion due to the unbalanced hybridization between V $3d$ metal and I $5p$ ligand states. Hybridization acts as an effective - although weak - trigonal crystal field, slightly lifting the degeneracy of the $t_{2g}^2$ orbitals in the ground state. The energy scale associated to the splitting underestimates the size of the experimental band gap, suggesting that the insulating ground state is stabilized by the electron correlation rather than via a JT distortion mechanism. The correspondence between experimental and simulated spectra, both for XMCD and XMLD, has been rarely reported in literature for vdW systems and for the first time in the case of VI$_3$. Our work paves the way for further explorations in parent systems, showing how ligand field theory is a powerful tool for describing the complex electronic structure of these compounds.


\section{Methods}
Measurements were performed at the XMCD endstation of the ID32 soft x-ray beamline at the European Synchrotron Radiation Facility (ESRF)\cite{brookes,kummer}. A commercial (\textit{HQGraphene}) bulk VI$_3$ crystal was mounted on a copper ID32-specific sample-holder using carbon tape and then exfoliated in N$_2$ flow inside the loadlock chamber connected to the ultrahigh vacuum (UHV) system, in order to expose a clean surface parallel to the (0001) crystallographic plane before the transfer to the measurement chamber. Due to the very hygroscopic nature of VI$_3$, both the sample preparation and the transfer towards the endstation were carried out under inert gas atmosphere, the first inside a glovebox, the second using an airtight transfer box.

The compound was studied in the 20-100 K temperature range covering the ferromagnetic (FM) and paramagnetic (PM) phases -- the magnetic transition temperature is $\sim51$ K -- under a background pressure lower than 3$\times$10$^{-10}$ mbar, by using right- and left-circularly as well as horizontal- and vertical-linearly polarized x-rays, depending on the specific technique. The synchrotron beam has a degree of linear polarization of almost 100\% and the monochromator has a resolving power better than 5000. The absorption signal was measured in total-electron yield (TEY) mode and normalized by the intensity collected on a gold mesh placed in front of the sample stage. The sample depth probed by XAS is estimated as 5 nm corresponding to 7 VI$_3$ layers. This amount of material is statistically significant to study the bulk properties of the compound.

Multiplet cluster calculations using the \textit{Quanty} code \cite{haverkort} have been carried out to simulate the spectra acquired at 20 K and to extract information about the FM ground state. The list of employed parameters include the octahedral crystal field ($10Dq$), an additional (trigonal) crystal field perturbation ($\tau$), intra-atomic interactions such as Hubbard ($U_{dd}$) and core-hole ($U_{pd}$) potentials, the $3d$ spin-orbit interaction ($\zeta$) and the charge-transfer energy ($\Delta$). The intra-atomic Coulomb and exchange interactions in the Hamiltonians have been calculated starting from Slater integrals reduced to 80\% of their Hartree-Fock values to account for screening effects. Hybridization between the V ion and I ligands is included by mixing the V $3d^2$ with the $3d^3\underline{L}$ configuration ($\underline{L}$ represents a hole on the adjacent ligands) by means of the hybridization parameters $V$ summing over states with corresponding symmetry. To simulate finite temperature effects at 20 K, ground state and excited states are averaged using Boltzmann statistics. Finally, the calculated spectra are convoluted with Lorentzian (0.5 eV) and Gaussian (0.4 eV) functions to account for the intrinsic lifetime and the instrumental broadening.


\section{\label{sec:level4}XNLD}
Linear dichroism in the soft X-rays is a powerful element specific technique to probe charge distribution asymmetries in transition metal compounds by means of linear polarization dependendent measurements. It allowed us to check the existence and the magnitude of the local distortion around the V atom and characterize the material electronic ground state. More in details, XNLD is the difference in the absorption spectra measured with two orthogonal linear polarizations in absence of any magnetic field (Figure \ref{fig:sketches}e). More specifically, in our experiment the beam was set 20$^\circ$ grazing on the sample surface. In this geometry, the electric field vectors for linear horizontal (LH) and linear vertical (LV) polarizations are respectively almost parallel and perpendicular to the [0001] direction ($c$ axis in Figure \ref{fig:sketches}d), enhancing the selective absorption along the out-of-plane and in-plane directions. The dichroic signal is thus interpreted as a consequence of the anisotropic charge density distribution in and out of the VI$_3$ plane, as due for instance to the elongation or compression of the VI$_6$ unit along the trigonal axis (Figure \ref{fig:sketches}c). A trigonal distortion in fact either elongates or contracts the pristine octahedral field along the three-fold axis, \textit{i.e.} the [111] direction in a cubic frame, and it represents a common deformation in vdW compounds with hexagonal or monoclinic symmetry \cite{Chang, sanjuan}. 

\begin{figure*}
    \centering
    \includegraphics[scale=0.90]{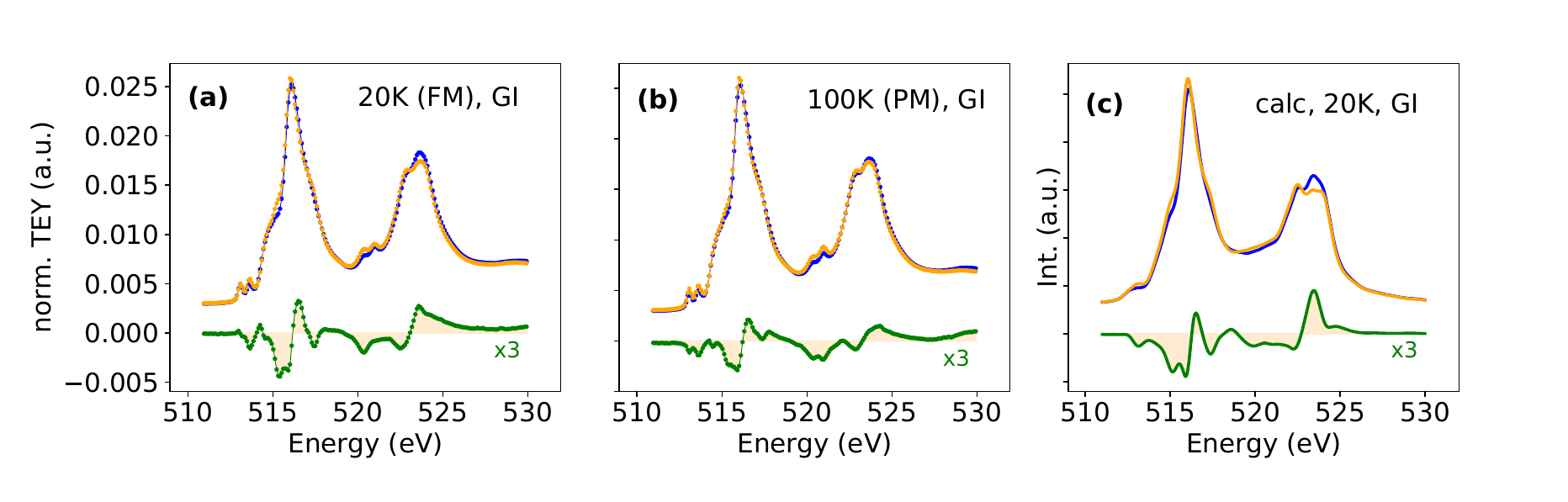}
    \caption{\textbf{XNLD on VI$_3$.} Experimental XAS and corresponding XNLD spectra of VI$_3$ measured across the V $L_{2,3}$ edges in the absence of any magnetic field and in grazing incidence (GI, 20$^\circ$) for the FM phase at 20 K (a) and the PM phase at 100 K, \textit{i.e.} above the structural transition (b). (c) Calculated XAS and XNLD spectra at 20 K. Experimental and theoretical data are reported with dotted and solid lines respectively. Horizontal- and vertical-linear polarization XAS and XNLD signal are shown respectively with blue, orange and green colors. Experimental XNLD signals are multiplied by a scale factor 3 for visualization.}
    \label{fig:xnld}
\end{figure*}

Figures \ref{fig:xnld}a-b show the XNLD spectra acquired at 20 and 100 K, \textit{i.e.} in the FM phase and in the PM phase above the structural transition temperature of VI$_3$ (79 K). The XAS are characterized by two pre-edge doublets at ~513.3 and ~520.6 eV and by two main peaks at 516.1 and 523.6 eV corresponding to the $L_3$ and $L_2$ absorption resonances due to the V $2p \rightarrow 3d$ electronic transition split by the core-hole spin-orbit interaction. The overall lineshape is strongly influenced by the multiplet structure given by crystal field and electron-electron interactions, notably Coulomb and exchange, which serves as a fingerprint for the inital state \cite{Vanderlaan1991,Vanderlaan2014}. The presence of a non zero - although weak - XNLD signal demonstrates that the VI$_6$ unit is not perfectly isotropic and its persistence both in the FM and PM phase indicates that its origin is not magnetic but related to a structural local distortion. However, the small size of the experimental XNLD signal ($\sim5\%$ of the average XAS intensity at the L$_3$) suggests that the perturbing crystal field acting on top of the octahedral coordination is weak and it leads to a modest splitting. The overall negative intensity -- we defined the XNLD signal as LH-LV -- is compatible with a higher electron charge density spread along the out-of-plane direction. The negative XNLD sign at the pre-edge doublets indicates that the latter arises from states with dominant in-plane character, while the positive features at the L$_3$ post edge (516.5 eV) and on the L$_2$ peak (523.6 eV) suggest instead that the highest energy occupied orbitals are oriented out-of-plane. Moreover, since the XNLD lineshape does not vary above the structural transition temperature of VI$_3$, we can conclude that the transition from monoclinic to rhombohedral phase modifies the stacking order of the neighbor VI$_3$ planes but not the intra-layer local V coordination and the orbital occupation.

We performed multiplet cluster calculations to simulate the absorption spectra acquired in the FM phase at 20 K for a trigonally distorted VI$_{6}$ unit by adopting a $D_{3d}$ point group symmetry, which represents a common octahedral ($O_h$) subgroup generated by breaking the pristine symmetry along the three-fold rotational axis. The parameter values that are best reproducing the experimental data have been listed in Table \ref{tab:table1}, while the simulated XAS and XNLD spectra are displayed in Figure \ref{fig:xnld}c. By setting the octahedral crystal field $10Dq=1.1$ eV, the charge-transfer energy $\Delta=3.0$ eV, and the Hubbard $U_{dd}=3.0$ eV, the obtained ground state for V is a hybridized mixture with 35\% $d^2$ and 65\% $d^3$ character, corresponding to a $d$-count of 2.65 e$^{-}/V$ atom, \textit{i.e.} more than the nominal total number of two electrons in an isolated V$^{3+}$ ion. This prominent amount of excess electrons is transferred from the I ligands and distributed on V states, giving VI$_3$ a covalent character. For comparison, we note that for CrI$_3$ a $d$-count of 3.41 electrons per Cr atom was extracted from the XAS/XMCD spectra, \textit{i.e.} more than the nominal value of three electrons of the Cr$^{3+}$ ion \cite{Frisk2018}. In transition metal compounds with moderate trigonal distortion, realistic values for the trigonal crystal field splitting are smaller than $10Dq$, and in the specific case of V$^{3+}$ compounds usually lower than 0.5 eV \cite{park}. In a recent work reporting theoretical XMCD calculations of VI$_3$ spectra, the trigonal crystal field was found equal to 0.15 eV \cite{hovancik}. In our simulations, the XNLD signal obtained by assigning $\tau$ values larger than 0.1 eV widely overestimates the experiments by a factor three. A correct size dichroism as the one reported in Figure \ref{fig:xnld}c has been obtained for $\tau=-0.003$ eV. Under this distortion, the estimated energy splitting of the degenerate $t_{2g}$ three-fold states corresponds to 2.8 meV. This value is higher than the thermal energy at 20 K.

\begin{table}[b]
\caption{\label{tab:table1}
Values of the main interactions acting within the VI$_6$ cluster unit used to calculate the XNLD spectra in Figure \ref{fig:xnld}c in ligand field approximation by adopting the $D_{3d}$ point group. $U_{dd}$ is the $3d$-$3d$ Coulomb potential, $U_{pd}$ is the $2p$-$3d$ core-hole potential, $10Dq$ is the $O_h$ crystal-field splitting, $\tau$ is the $D_{3d}$ crystal-field splitting, $V_{e_g}$, $V_{a1_{g}}$, and $V_{e^{\prime}_g}$ are the hybridization parameters for each of the given $D_{3d}$ symmetry sub-groups, $\zeta_{i,f}$ is the $3d$ spin-orbit interaction (initial/final state), and $\Delta$ is the charge-transfer energy. Values are expressed in eV.}
\begin{indented}
\item[]\begin{tabular}{lllllllllll}
\hline \hline
& $U_{dd}$ & $U_{pd}$ & $10Dq$ & $\tau$ & $V_{e_g}$ & $V_{a_{1g}}$ & $V_{e^{\prime}_g}$ & $\zeta_i$ & $\zeta_f$ & $\Delta$ \\
\hline \hline
$e_g^{\prime}a_{1g}$ & 3.000 & 4.290 & 1.100 & -0.003 & 1.700 & 1.100 & 1.100 & 0.027 & 0.036 & 3.000 \\
\hline \hline
\end{tabular}
\end{indented}
\end{table}

A deeper understanding of the origin of the observed local distortion is however achieved by properly taking into account the hybridization between the V $3d$ and I $5p$ ligand orbitals. For a $3d$ electron system described within the ligand theory framework, the trigonal distortion is mainly caused by the ligand atoms and the effects are mostly reflected in the hybridization parameters. Hence, the same simulation output shown in Figure \ref{fig:xnld}c can be obtained by neglecting the trigonal field, \textit{i.e.} setting $\tau=0$, and by properly modeling the hybridization potentials.

\begin{figure*}
    \centering
    \includegraphics[scale=0.85]{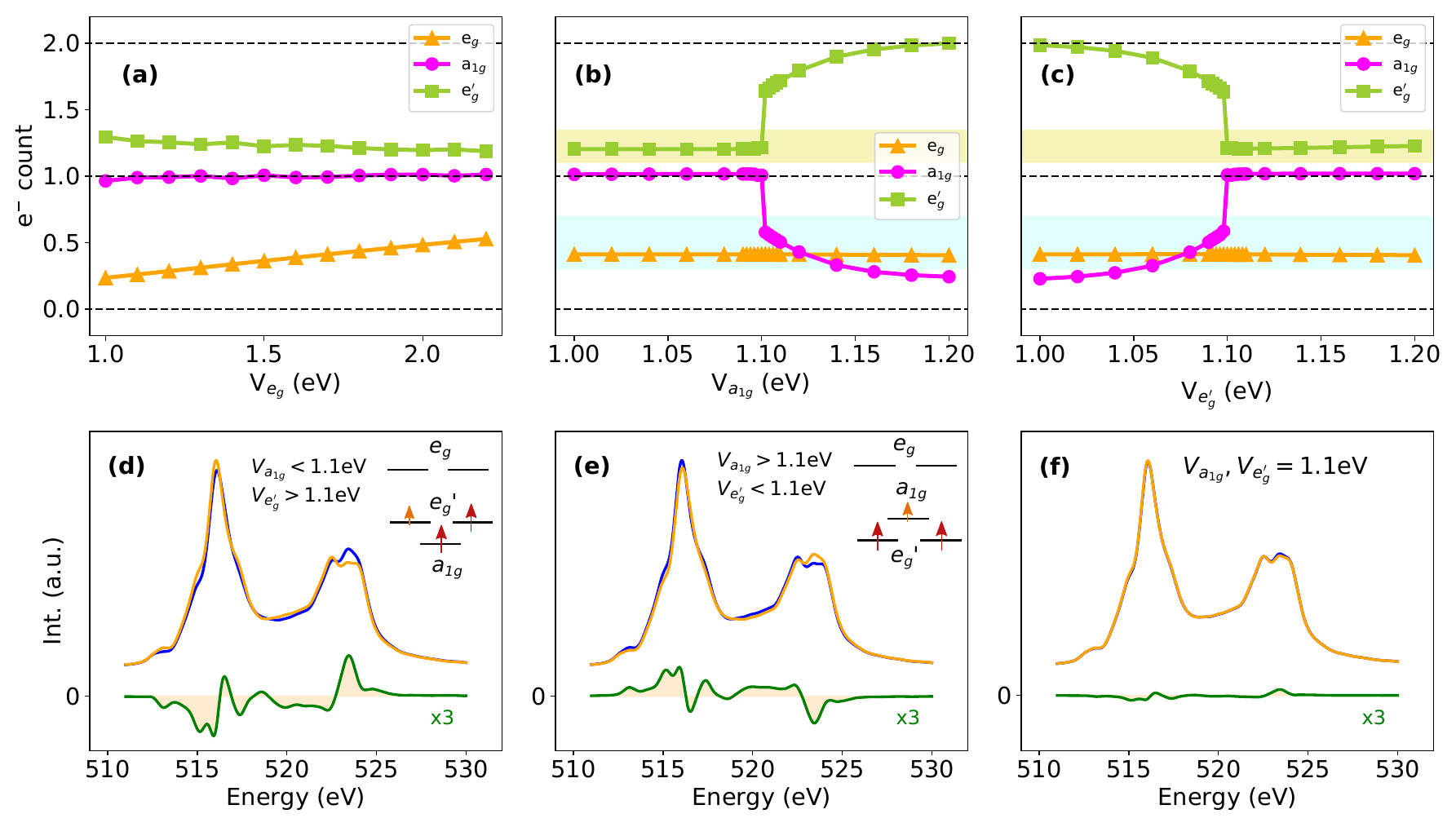}
    \caption{\textbf{Effects of hybridization on XNLD spectra.} (a-c) Calculated orbital occupations of the $e_{g}$ (orange triangles), $a_{1g}$ (pink circles) and $e_g^{\prime }$ (green squares) states per V atom as function of the symmetry adapted potentials $V_{e_g}$, $V_{a_{1g}}$ and $V_{e_g^\prime}$. At each time one of the potential was varied by keeping the others constant to the default values in Table \ref{tab:table1}. Ochre and cyan areas denote excess electrons on $e_g^\prime$ and $a_{1g}$ states respectively. (d-f) Calculated XAS and XNLD spectra at 20 K and 20$^\circ$ grazing incidence for $\tau=0$ eV and for $V_{a_{1g}}=1.098$ eV, $V_{e_g^\prime}=1.102$ eV (d), $V_{a_{1g}}=1.102$ eV, $V_{e_g^\prime}=1.098$ eV (e) and $V_{a_{1g}}=V_{e_g^\prime}=1.1$ eV (e). The schematics in the insets illustrate the corresponding state configuration and electronic occupation. Long red and short orange arrows correspond to integer and fractional electron count on V $3d$ states. The XAS spectra calculated with horizontal and vertical polarization are shown in blue and orange, while the XNLD signal is green.}
    \label{fig:xnld_hybrid}
\end{figure*}

In Figure \ref{fig:xnld_hybrid}, we studied the effects of the hybridization parameters on the sign and lineshape of the calculated XNLD signal. Figures \ref{fig:xnld_hybrid}a-c display how the electronic occupation varies within the V $d$-states as a function of the symmetry adapted potentials $V_{e_g}$, $V_{a_{1g}}$ and $V_{e_g^{\prime}}$. The $e_g^\prime$ and $a_{1g}$ states descend directly from the $t_{2g}$ irreducible representation upon the $O_h$-to-$D_{3d}$ symmetry lowering, while the $e_g$ states branch to the same irreducible representation in $O_h$ and $D_{3d}$ point groups and their are not relevant to the distortion. We can observe that, while the electron count within the $e_g$ state (orange triangles) is exclusively due to the charge transfer from the ligands and it remains nearly constant at 0.4 e$^{-}/V$ atom for all the three cases, the electron content within the $e_g^\prime$ (green squares) and $a_{1g}$ (pink circles) states is mainly related to V $3d$-electrons modulated by the hybridization with the I ligands, and it varies symmetrically as function of the $V_{a_{1g}}$ and $V_{e_g^\prime}$ potentials (Figures \ref{fig:xnld_hybrid}b-c). By reducing (increasing) the hybridization parameter corresponding to the $a_{1g}$ ($e_g^\prime$) symmetries below (above) the threshold value of 1.1 eV, an extra fractional number of electrons is registered on the $e_g^\prime$ orbitals (electron count higher than 1, plot area highlighted in ochre). Viceversa, by increasing (reducing) the hybridization potential $V_{a_{1g}}$ ($V_{e_g^\prime}$) above (below) 1.1 eV, a fractional electron excess would be localized on the $a_{1g}$ orbitals (electron count higher than 0, plot area highlighted in cyan). This extra orbital occupation is due to electrons transferred from the I ligands to V orbitals with specific in-plane or out-of-plane character. They change the V $d$-shell charge density distribution from spherical ($O_h$) to slightly anisotropic, according to the shape of the orbitals accommodating the transferred electrons. From the spectroscopic point of view, it results into a difference between the absorption by horizontal and vertical linearly polarized light, as shown by the simulated spectra in Figures \ref{fig:xnld_hybrid}d-e. In particular, for the two described cases, the net dichroism sign is inverted. For $V_{a_{1g}}$ and $V_{eg^{\prime}}$ both equal to 1.1 eV, the XNLD signal remains zero and the extra charge is equally spread among the orbitals, leading to isotropic charge distribution around the V ions (Figure \ref{fig:xnld_hybrid}f).

As illustrated in the insets of Figure \ref{fig:xnld_hybrid}d-e, the orbital occupancy is the consequence of the stabilization of two different ground state configurations, \textit{i.e.} ($e_g^{\prime},a_{1g}$) and $e_g^{\prime}\,^2$ \cite{park,nguyen,yang}. When the $V_{a_{1g}}$ and $V_{e_g^\prime}$ hybridization parameters are respectively above and below 1.1 eV, the orbitals are filled with two electrons lying in the doubly degenerate $e_g^{\prime}$ state, whereas the $a_{1g}$ level higher in energy remains vacant (Figure \ref{fig:xnld_hybrid}e). Although this orbital occupation is compatible with an insulator, the calculated XNLD signal has overall opposite sign compared to the experiment (Figure \ref{fig:xnld}a). The excess electrons transferred from the I ligands to the V atoms (+0.25$e^-$) are localized on the $a_{1g}$ state, which has higher out-of-plane character. On the other hand, for $V_{a_{1g}}$ and $V_{e_g^\prime}$ values below and above 1.1 eV, excellent agreement between theoretical and experimental white lines is achieved (Figures \ref{fig:xnld_hybrid}d). In this case, the hybridization flips the split energy levels in the opposite way, with the $e_g^{\prime}$ states higher in energy than the $a_{1g}$ state. In contrast to the previous case, a similar excess of electrons lies instead within the $e_g^{\prime}$ states having more in-plane character. This $e_g^{\prime}a_{1g}$ configuration corresponds to a metallic ground state. However, in strongly correlated materials this configuration might still have an insulating ground state for large enough Coulomb interaction \cite{georges}, especially in the case of a weak distortion.

To achieve a suitable quantitative agreement between the theoretical and experimental XNLD intensities, the mismatch between $V_{a_{1g}}$ and $V_{e_g^\prime}$ has to be small. We found that to reach a dichroism as low as $\sim5\%$ (estimated at the L$_3$ edge), $V_{a_{1g}}$ and $V_{e_g^\prime}$ have to be reduced (increased) by about 2 meV from the zero XNLD value, \textit{i.e.} 1.1 eV (the full set of parameter values used to calculate the spectra in Figures \ref{fig:xnld_hybrid}d-f is reported in Table \ref{tab:table2}. The energy gain of the ground state associated to such $e_g^{\prime}a_{1g}$ configuration lies $\sim2.2$ meV below the unsplit $t_{2g}^2$ case. The energy splitting of the degenerate $t_{2g}^2$ as extracted from the calculations is 2.8 meV, higher than the thermal energy at 20K but too low to account for the half eV band gap as found experimentally. This result, obtained by properly tuning the hybridization parameters represents a description of the distortion in VI$_3$ equivalent to apply a negative few meV high trigonal crystal field interaction. In fact, within a molecular orbital picture, hybridization potentials of a specific symmetry lift the orbital degeneracy by pushing the corresponding bonding-antibonding states away in energy. In this way, the hybridization acts as an effective (trigonal) crystal field resulting in the splitting of the $t_{2g}$ orbitals (shown in the insets of Figure \ref{fig:xnld_hybrid}) and in the compression/elongation of the atomic bonding along the trigonal axis. We conclude that the linear dichroism emerges from an unbalanced charge transfer of electrons from $5p$ I ligand towards the $3d$ V metal states with corresponding symmetry and it establishes hybridization as the source of the trigonal distortion in the VI$_3$ system.

\begin{table}[b]
\caption{\label{tab:table2}
Trigonal crystal field and hybridization potentials as used to calculate the XNLD spectra in Figures \ref{fig:xnld_hybrid}d-f and the XMCD spectra in Figure \ref{fig:xmcd}c for each of the three electronic configurations described in the text: $e_g^{\prime}a_{1g}$, $e_g^{\prime}\,^2$,  $t_{2g}^2$. For all the other parameters in the calculations, the values are those reported in Table \ref{tab:table1}. Values are expressed in eV.}
\begin{indented}
\item[]
\begin{tabular}{lllllllllll}
\hline \hline
& $\tau$ & $V_{e_g}$ & $V_{a_{1g}}$ & $V_{e^{\prime}_g}$ \\
\hline \hline
$e_g^{\prime}a_{1g}$ & 0.000 & 1.700 & 1.098 & 1.102 \\
\hline
$e_g^{\prime}\,^2$ & 0.000 & 1.700 & 1.102 & 1.098 \\
\hline
$t_{2g}$ & 0.000 & 1.700 & 1.100 & 1.100 \\
\hline \hline
\end{tabular}
\end{indented}
\end{table}

It turns out that the trigonal distortion appears as a concomitant but minor contribution to the insulating band gap opening. VI$_3$ $3d$ electrons have in fact relatively localized nature -- as shown by the corresponding low dispersing valence bands in Ref\cite{devita} -- and they are only weakly sensitive to structural distortions. We deduce that the $t_{2g}$ splitting is instead dominated by the strength of Coulomb and exchange energies, which are typically of the order of several eV (such interactions are controlled in our calculations by the $U_{dd}$ and $U_{pd}$ parameters reported in Table \ref{tab:table2}). However, the lifted degeneracy caused by the distortion can in principle reduce the Coulomb interaction in the system\cite{georges}. In conclusion, although JT effect and electron correlation are not two mutually exclusive mechanisms, our results strengthen the hypothesis of the VI$_3$ insulating ground state stabilized mainly by intra-atomic electron-electron interactions rather than by symmetry lowering due to JT distortion, in agreement with the conclusions from previous photoemission studies\cite{devita}.


\section{XMCD}
A confirmation of these interpretations can be obtained by calculating also XMCD spectra starting from the same set of parameters optimized for the XNLD simulations. XMCD is the difference between two XAS spectra acquired with left- ($\sigma^{-}$) and right- ($\sigma^{+}$) circularly polarized x-rays with the magnetic field applied parallel to the beam direction (Figure \ref{fig:sketches}f). At V L$_{2,3}$ absorption edges, opposite beam helicities, \textit{i.e.} parallel and antiparallel to the \textit{2p} electron orbital moments, excites electrons with opposite spins. The difference between the number of available spin-up and spin-down hole states turns out into a net negative or positive peak. 

\begin{figure*}
    \centering
    \includegraphics[scale=0.90]{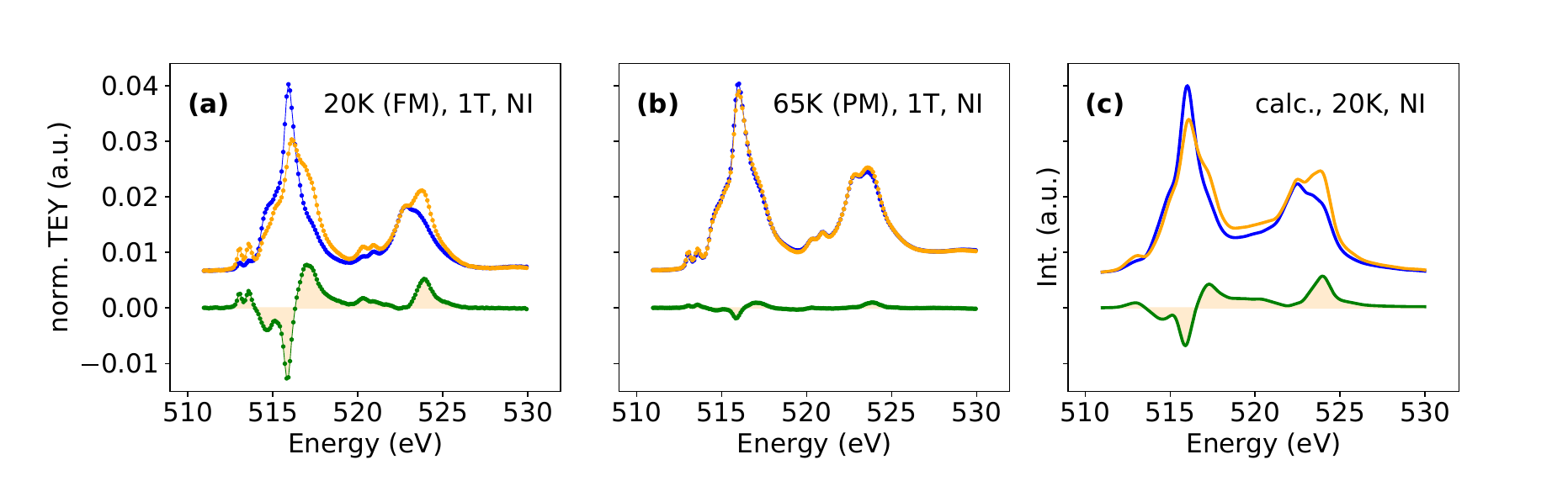}
    \caption{\textbf{XMCD on VI$_3$.} Experimental XAS and corresponding XMCD spectra of VI$_3$ across the V $L_{2,3}$ edges measured at 1 T magnetic field and in normal incidence (NI): (a) FM phase (20 K) and (b) PM phase (65 K). (c) Theoretical XAS and XMCD spectra calculated at 20 K. Experimental and theoretical data are reported with dotted and solid lines respectively. Left- and right-circular polarization XAS and XNLD signals are shown respectively with blue, orange and green colors.}
    \label{fig:xmcd}
\end{figure*}

Figures \ref{fig:xmcd}a,b show the VI$_3$ XMCD spectra measured at 20 and 65 K, \textit{i.e.} in the FM and PM phases respectively. The spectra were acquired with the sample (0001) plane normal to the incident beam and by applying a magnetic field of 1 T parallel to the beam direction, sufficient to saturate all the magnetic domains \cite{tian}. Also in this case, the XAS are characterized by two main peaks at 516.1 and 523.6 eV corresponding to the $L_3$ and $L_2$ absorption resonances and by two pre-edge doublets at ~513.3 and ~520.6 eV. Both in the L$_3$ and L$_2$ regions, right- and left-circularly polarized light generate different XAS spectra at 20 K, giving rise to a large dichroism, estimated $\sim$45\% of the average XAS intensity at the L$_3$ edge. The dichroic signal is almost negligible instead at 65 K, as expected in the PM phase.

As the same for XNLD, we performed multiplet cluster calculations to simulate the absorption spectra acquired in the FM phase (20 K) using $D_{3d}$ point group symmetry and the parameters reported in Table \ref{tab:table2} corresponding to the $e_g^{\prime}a_{1g}$ configuration. The XAS white lines for left and right circular polarization and the XMCD signal reproduce qualitatively well the experimental data. However they underestimate the XMCD dicroism negative peak at L$_3$. Again, the obtained ground state for V is a hybridized mixture with 35\% $d^2$ and 65\% $d^3$ character, and it is due to a prominent amount of electrons transferred from the $5p$ I ligands on the $3d$ V states, giving VI$_3$ a covalent character. 
As calculated VI$_3$ has a high spin character, with spin moment $S=-0.99\mu_B$ at 0 K, shrinking to $S=-0.38\mu_B$ at 20 K, and orbital moment $L=1.18\mu_B$ at 0 K, shrinking to $L=0.44\mu_B$ at 20 K. In particular, the high obtained orbital moment at 0 K matches the one calculated in Ref\cite{hovancik}, and it is considered as the responsible of the high magnetocrystalline anisotropy observed in VI$_3$ \cite{tian}. The general good agreement between experimental and theoretical XAS obtained also for the XMCD data establishes the robustness of our model and interpretations discussed in the XNLD section and it demonstrates the good consistency between the information extracted from the two data sets.


\section{Conclusions}
In summary, experimental XNLD spectra of VI$_3$ bulk crystals below and above magnetic and structural transitions displaying finite dichroism signal show that the charge distribution around the V atoms is slightly anisotropic. Such anisotropy is theoretically accounted for by comparing the experimental data with multiplet cluster calculations, more specifically by properly including hybridization between metal and ligand states, which acts as an effective trigonal crystal field distorting the VI$_6$ unit. We deduced that the degeneracy in the ground state VI$_3$ high-spin $t_{2g}^2$ configuration is slightly lifted by metal-ligand orbital hybridization causing a weak but finite trigonal distortion. Theoretical agreement with experimental data has been obtained both for XMCD and XNLD using the same set of parameters, demonstrating the consistency between the magnetic and orbital information extracted from the two cases. However, the estimated $t_{2g}$ levels splitting of few meV resulting into the measured XNLD signal is far too low to account for the half eV band gap observed in previous experimental works. We suggest that the insulating character of VI$_3$ emerges instead from electron correlation effects, notably Coulomb and exchange interactions. These conclusions are in agreement with previous photoemission studies. Our results represent both a theoretical and experimental benchmark for the interpretation of the XAS spectra and the fundamental properties of other transition metal halides, \textit{e.g.} CrI$_3$, and oxides, but also in comparison to other oxide forms of VI$_3$ crystals synthesized as single or few layer materials, whose properties might be altered by the interface proximity or reduced dimensionality.


\ack{Acknowledgments}
We acknowledge the European Synchrotron Radiation Facility and ELETTRA Sincrotrone Trieste for provision of beamtime on the ID32 and on the APE-HE and LE beamlines. This work has been partially performed in the framework of the Nanoscience Foundry and Fine Analysis (NFFA-MUR Italy Progetti Internazionali) facility (https://www.trieste.nffa.eu/).
The authors thank Dr. D. Betto and Dr. K. Kummer of the ID32 beamline at ESRF for the fruitful discussions on the use of \textit{Quanty}.


\section*{References}
\bibliographystyle{iopart-num}
\bibliography{biblio}

\providecommand{\noopsort}[1]{}\providecommand{\singleletter}[1]{#1}%
\providecommand{\newblock}{}
\begin{thebibliography}{10}
\expandafter\ifx\csname url\endcsname\relax
  \def\url#1{{\tt #1}}\fi
\expandafter\ifx\csname urlprefix\endcsname\relax\def\urlprefix{URL }\fi
\providecommand{\eprint}[2][]{\url{#2}}

\bibitem{huang}
Huang B, Clark G, Navarro-Moratalla E, Klein D~R, Cheng R, Seyler K~L, Zhong D,
  Schmidgall E, McGuire M~A, Cobden D~H {\em et~al.\/} 2017 {\em Nature\/} {\bf
  546} 270--273

\bibitem{zaanen}
Zaanen J, Sawatzky G and Allen J 1985 {\em Phys. Rev. L\/} {\bf 55} 418

\bibitem{degroot2}
De~Groot F~M, Fuggle J, Thole B and Sawatzky G 1990 {\em Physical Review B\/}
  {\bf 42} 5459

\bibitem{degroot1}
De~Groot F 2005 {\em Coordination Chemistry Reviews\/} {\bf 249} 31--63

\bibitem{bora}
Bora M and Deb P 2021 {\em Journal of Physics: Materials\/} {\bf 4} 034014

\bibitem{zhao}
Zhao G~D, Liu X, Hu T, Jia F, Cui Y, Wu W, Whangbo M~H and Ren W 2021 {\em
  Physical Review B\/} {\bf 103} 014438

\bibitem{zhangVSe2}
Zhang W, Zhang L, Wong P~K~J, Yuan J, Vinai G, Torelli P, van~der Laan G, Feng
  Y~P and Wee A~T 2019 {\em ACS \ce{n}ano\/} {\bf 13} 8997--9004

\bibitem{vinai}
Vinai G, Bigi C, Rajan A, Watson M~D, Lee T~L, Mazzola F, Modesti S, Barua S,
  Hatnean M~C, Balakrishnan G {\em et~al.\/} 2020 {\em Physical Review B\/}
  {\bf 101} 035404

\bibitem{he}
He J, Ma S, Lyu P and Nachtigall P 2016 {\em Journal of Materials Chemistry
  C\/} {\bf 4} 2518--2526

\bibitem{long}
Long C, Wang T, Jin H, Wang H and Dai Y 2020 {\em The Journal of Physical
  Chemistry Letters\/} {\bf 11} 2158--2164

\bibitem{tian}
Tian S, Zhang J~F, Li C, Ying T, Li S, Zhang X, Liu K and Lei H 2019 {\em
  Journal of the American Chemical Society\/} {\bf 141} 5326--5333

\bibitem{nguyen}
Nguyen T~P~T, Yamauchi K, Oguchi T, Amoroso D and Picozzi S 2021 {\em Physical
  Review B\/} {\bf 104} 014414

\bibitem{mcguire}
McGuire M~A, Dixit H, Cooper V~R and Sales B~C 2015 {\em Chemistry of
  Materials\/} {\bf 27} 612--620

\bibitem{dolevzal}
Dole{\v{z}}al P, Kratochv{\'\i}lov{\'a} M, Hol{\`y} V, {\v{C}}erm{\'a}k P,
  Sechovsk{\`y} V, Du{\v{s}}ek M, M{\'\i}{\v{s}}ek M, Chakraborty T, Noda Y,
  Son S {\em et~al.\/} 2019 {\em Physical Review Materials\/} {\bf 3} 121401

\bibitem{kong}
Kong T, Stolze K, Timmons E~I, Tao J, Ni D, Guo S, Yang Z, Prozorov R and Cava
  R~J 2019 {\em Advanced Materials\/} {\bf 31} 1808074

\bibitem{son}
Son S, Coak M~J, Lee N, Kim J, Kim T~Y, Hamidov H, Cho H, Liu C, Jarvis D~M,
  Brown P~A {\em et~al.\/} 2019 {\em Physical Review B\/} {\bf 99} 041402

\bibitem{kundu}
Kundu A~K, Liu Y, Petrovic C and Valla T 2020 {\em Scientific reports\/} {\bf
  10} 1--8

\bibitem{devita}
De~Vita A, Nguyen T~T~P, Sant R, Pierantozzi G~M, Amoroso D, Bigi C, Polewczyk
  V, Vinai G, Nguyen L~T, Kong T {\em et~al.\/} 2022 {\em Nano Letters\/}

\bibitem{park}
Park J~H, Tjeng L, Tanaka A, Allen J, Chen C, Metcalf P, Honig J, De~Groot F
  and Sawatzky G 2000 {\em Physical Review B\/} {\bf 61} 11506

\bibitem{yang}
Yang K, Fan F, Wang H, Khomskii D and Wu H 2020 {\em Physical Review B\/} {\bf
  101} 100402

\bibitem{brookes}
Brookes N~B, Yakhou-Harris F, Kummer K, Fondacaro A, Cezar J, Betto D,
  Velez-Fort E, Amorese A, Ghiringhelli G, Braicovich L {\em et~al.\/} 2018
  {\em Nuclear Instruments and Methods in Physics Research Section A:
  Accelerators, Spectrometers, Detectors and Associated Equipment\/} {\bf 903}
  175--192

\bibitem{kummer}
Kummer K, Fondacaro A, Jimenez E, Velez-Fort E, Amorese A, Aspbury M,
  Yakhou-Harris F, Van Der~Linden P and Brookes N 2016 {\em Journal of
  synchrotron radiation\/} {\bf 23} 464--473

\bibitem{haverkort}
Haverkort M~W 2016 {\em Journal of Physics: Conference Series\/} {\bf 712}
  012001

\bibitem{Chang}
Chang A~G, Lan L~W, Chan Y~J, Kuo C~N, Chen T, Huang C~H, Chuang T~H, Wei D~H,
  Lue C~S and Kuo C~C 2022 {\em Physical Review B\/} {\bf 106} 125412

\bibitem{sanjuan}
Sanju{\'a}n M, Kanehisa M and Jouanne M 1992 {\em Physical Review B\/} {\bf 46}
  11501

\bibitem{Vanderlaan1991}
{van der Laan} G and Thole B~T 1991 {\em Physical Review Letter\/} {\bf 43}
  13401 -- 13411

\bibitem{Vanderlaan2014}
{van der Laan} G and Figueroa A~I 2014 {\em Coord. Chem. Rev.\/} {\bf 277-278}
  95--129

\bibitem{Frisk2018}
Frisk A, Duffy L~B, Zhang S, {van der Laan} G and Hesjedal T 2018 {\em
  Materials Letters\/} {\bf 231} 5

\bibitem{hovancik}
Hovancik D, Pospisil J, Carva K, Sechovsky V and Piamonteze C 2023 {\em Nano
  Letters\/} {\bf 23} 1175--1180

\bibitem{georges}
Georges A, Medici L~D and Mravlje J 2013 {\em Annu. Rev. Condens. Matter
  Phys.\/} {\bf 4} 137--178

\end{thebibliography}
\clearpage


\clearpage

\clearpage

\clearpage


\end{document}